\documentclass{iopart}

\usepackage{array}
\usepackage{graphicx}
\usepackage{bigstrut}

\usepackage{color}

\graphicspath{{fig/}}

\newcommand{\ket}[1]{\left.\left|{#1} \right\rangle\right.}
\newcommand{\squ}{\hspace*{-0.1cm}}

\begin{document}

\title{Towards Mott design by $\delta$-doping of strongly correlated titanates}

\author{Frank Lechermann and Michael Obermeyer}
\address{I. Institut f{\"u}r Theoretische Physik, Universit{\"a}t Hamburg, 
D-20355 Hamburg, Germany}

\begin{abstract}
Doping the distorted-perovskite Mott insulators LaTiO$_3$ and GdTiO$_3$ with a single SrO layer
along the [001] direction gives rise to a rich correlated electronic structure. 
A realistic superlattice study by means of the charge self-consistent combination of density 
functional theory with dynamical mean-field theory reveals layer- and temperature-dependent 
multi-orbital metal-insulator transitions. An orbital-selective metallic layer at the interface 
dissolves via an orbital-polarized doped-Mott state into an orbital-ordered insulating regime 
beyond the two conducting TiO$_2$ layers. 
We find large differences in the scattering behavior within the latter. Breaking 
the spin symmetry in $\delta$-doped GdTiO$_3$ results in blocks of ferromagnetic itinerant and  
ferromagnetic Mott-insulating layers which are coupled antiferromagnetically.
\end{abstract}

%\pacs{73.20.-r,71.27.+a, 75.70.Cn}
\maketitle

\section{Introduction}
In view of future technological applications, the investigation of oxide heterostructures provides 
the possibility for exploring novel composite materials beyond nature's original conception 
(see e.g.~\cite{zub11,hwa12,cha14}) for reviews). Additionally, this research extends 
the concept of materials design towards the realm of strongly correlated
systems. Many of the essential heterostructure building blocks either harbor 
partially-filled transition-metal $d$-shells already in their bulk configuration
(e.g. LaTiO$_3$, LaVO$_3$, GdTiO$_3$, etc.) or display such partial filling due to
interface doping (e.g. for SrTiO$_3$). Therefore weakly-screened Coulomb interactions may 
give rise to explicit many-body effects that eventually govern the materials physics.
Unusual metallicity from combining bulk band insulators, emergent magnetic order, 
superconductivity or large thermopower are only a few fascinating phenomena that occur in the 
interfacing regions. 
Engineering such structured matter allows for a direct manipulation of systems in
or close to the Mott-critical regime, i.e. offers the potential for {\sl Mott design}.
As a specific realization thereof, the $\delta$-doped oxide 
heterostructures~\cite{kim10,jan11,cho12,ras14}, i.e. introducing well-defined impurity 
monolayers into a given host oxide compound, recently emerged as a canonical method to create 
challenging electronic states from experimental fabrication. Especially the $\delta$-doping of Mott 
insulators~\cite{oue13,che13,lec13,jac14} not only has relevance in the designing context, but 
furthermore sheds light on the generic physics of the realistic doped-Mott state in a controlled 
way without the usual complications arising from disorder and other features of random impurity 
doping.

In this work we use a first-principles many-body approach to investigate the $\delta$-doping of
the Mott insulators LaTiO$_3$ (LTO) and GdTiO$_3$ (GTO). The bulk materials are examples of two 
qualitatively different trends within the series of perovskite-like $R$TiO$_3$ compounds 
($R$=rare-earth element)~\cite{kom07} with GdFeO$_3$-type distortion. 
While for smaller rare-earth ionic radius (Yb, Y, Gd) the systems become ferromagnetic at low 
temperature, for larger radius (Sm, Nd, La) the materials are $G$-type antiferromagnetic in the 
crystallographic $a$ direction upon cooling. For GTO the Curie temperature is 36K and the
N{\'e}el temperature of LTO amounts to 146K. An increasing deviation of the Ti-O-Ti bond angle 
from 180$^{\circ}$ from La to Y is key to the change of magnetic ground state among the 
titanates~\cite{gor83,amo96,ono97,hay99,kei00}. The crucial low-energy electron states are 
dominated by the threefold of Ti($t_{2g}$) orbitals, with a nominal filling of one electron.
Numerous theoretical studies are devoted to reveal the electronic structure of these bulk Mott 
insulators~\cite{kha00,sol04,mos01,pav04,cra04,pav05,ior12}. Especially the very detailed ab-inito 
study of Pavarini {\sl et. al}~\cite{pav05} covers many aspects of the representants LaTiO$_3$ and 
YTiO$_3$, from the generic band structure to crystal-field/tight-binding considerations, up to the 
inclusion of correlation effects. The latter are shown to be important in driving significant 
orbital ordering in these $3d^{1}$ materials, which renders other orbital-degenerate modelings 
questionable.

Main intention here is to shed light on the correlation physics of $\delta$-doped heterostructures
as well as to promote the $R$TiO$_3$ physics to the next level by investigating the behavior
from doping with a single SrO layer. We thereby
concentrate on the temperature regimes above the bulk magnetic ordering. Two main question arise
in the latter scenario: the first one concerncs the appearance of metallization via the non-random 
hole doping, and if it sets in, the characterization in terms of the multi-orbital Ti manifold and
its layer dependence. When a (layer-dependent) Mott state is reached, its possible 
deviation from the known bulk Mott-insulating state is secondly of key interest. The 
clarification of the correlated electronic structure in this canonical structure case is highly
relevant for the understanding of other layerings and possible applications, e.g. in the
area of solar cells~\cite{ass13}.

Using the charge self-consistent combination of density functional theory (DFT) with 
dynamical mean-field theory (DMFT) in a superlattice architecture we indeed reveal a very rich
electronic-structure phenomenology. Layer-dependent multi-orbital metal-insulator transitions via 
$\delta$-doping are identified for the case of SrO/LTO and SrO/GTO heterostructures, that involve 
three different electronic regimes. With increasing distance to the doping layer, an $xy$-dominated 
metallic state settling right at the interface is replaced by an orbital-polarized doped-Mott 
metallic layer which transforms to bulk-modified insulating Mott layers. Embedding such
complex electron states within oxide heterostructures opens the door for many ways of manipulations 
and can lead to the identification of building blocks for new devices.

\section{Theoretical framework}
On the level of the local density approximation (LDA) to density functional theory (DFT), a 
mixed-basis pseudopotential framework~\cite{lou79,mbpp_code} is applied for the 
structural optimization of the $\delta$-doped supercells and the basic
electronic structure investigation. Norm-conserving pseudpotentials as well as a combined basis
of plane waves and localized functions for Ti$(3d)$ and O$(2s2p)$ is utilized. The latter
allow for a moderate plane-wave cutoff energy $E_{\rm cut}$=13Ryd for the demanding
supercell calculations, which are based on a 5$\times$5$\times$3 $k$-point mesh in reciprocal space.
Though $f$-electrons can be also treated also in many-body bulk computations, we here aim for novel 
correlations effects dominantly from first-principles $3d$ electrons in challenging heterostructure 
architectures. Thus effects of possible $4f$ states are captured approximately. In La those states 
are unoccupied in the atom and form empty conduction states in LTO. One may neglect them for the 
present purpose to a good approximation in the pseudopotential construction and the 
localized-function basis. 
Contrary to the neighboring band-insulating EuTiO$_3$ compound with Eu$^{2+}$ and Ti$^{4+}$, 
GTO harbors Gd$^{3+}$ and Ti$^{3+}$. This shifts the half-filled $4f$ shell (occupation 7 
electrons) shell of Gd further down in energy and with an assumed large intra-orbital Coulomb 
interaction this local mainfold, deep in energy, has no quasiparticle contribution at low energy. 
Hence the Gd($4f$) shell is not of eminent importance in GdTiO$_3$ and is here put in the 
pseudopotential frozen core.

Full local electronic correlation effects beyond the static (i.e. LDA, LDA+U) realm 
are taken care of by the dynamical mean-field theory (DMFT) within the framework of 
charge self-consistent 
DFT+DMFT~\cite{sav01,pou07,gri12}. The multi-site and -orbital correlated subspace where 
explicit Coulomb interactions are treated on the many-body level is here defined by the
$t_{2g}$-like low-energy crystal-field (cf) bases of the various Ti ions in the given structures. It
is obtained by tailored projected local orbitals~\cite{ani05,ama08,hau10} based on a set of 
low-energy Kohn-Sham (KS) states. The orbital projections are given
by linear combinations of the original ($xz$, $yz$, $xy$) functions that diagonalize the local
$t_{2g}$-like 3$\times$3 orbital density matrix on each Ti ion, respectively. In case of the bulk 
compounds, the 12 low-energy KS bands close to the Fermi level are used to facilitate the projection.
All Ti ions are equivalent by symmetry in the bulk GdFeO$_3$ structure and the 
site-dependent projections can be chosen such that a single 3$\times$3 self-energy matrix 
${\bf \Sigma}(\omega)$ is converged. For the $\delta$-doped compounds, a multi-site DFT+DMFT 
scheme is put into practise with 5 inequivalent Ti ions in the chosen supercells 
(see sections~\ref{sec:struc} and~\ref{sec:dftdmft-delta}). There a KS manifold of 60 low-energy
bands enters the projection. 

A rotational-invariant three-orbital Hubbard Hamiltonian in Slater-Kanamori paramterization is 
applied on each Ti site. The intra-orbital Coulomb interaction is chosen as $U$=5eV and the 
Hund's exchange as $J_{\rm H}$=0.64eV, identical to former LDA+DMFT studies of bulk 
titanates~\cite{pav04,pav05}. Continuous-time quantum Monte Carlo in the hybridization 
expansion~\cite{rub05,wer06,triqs_code,boe11} is utilized to solve the DMFT impurity problems. 
At each correlated Ti site $i$ and orbital $m$ a double-counting (DC) correction of the fully-localized 
form~\cite{sol94} making use of the local particle-number operator $n_i$, 
i.e. $\Sigma^{\rm DC}_{im\sigma}$=$U(\langle n_i\rangle$$-$$1/2)
$$-$$J_{\rm H}(\langle n_{i\sigma}\rangle$$-$$1/2)$ with $\sigma$=$\uparrow$,$\downarrow$, 
is applied to the self-energy in the complete charge self-consistent convergency cycle. 
The resulting local and total spectral functions are derived from analytical continuation 
of the Green's functions in Matsubara space via the maximum-entropy method. 

Note that different or larger correlated subspaces, e.g. by including 
O$(2p)$-dominated KS band states in the projection, are possible and currently under
investigation for oxide bulk Mott insulators~\cite{hau14,dan14}. But since our focus is
on the challenging large-scale heterostructure problem where such recent extensions are very 
expensive, we work in minimal $t_{2g}$-based correlation framework.
%%%%%%%%%%%%%%%%%%%%%%%%%%%%%%%%%%%%%%%%%%%%%%%%%%%%%%%%%%%%%%%%%%%%%%%%%%%%%%%%%%%%%%%%%
\begin{figure}[t]
\begin{center}
\hspace*{-0.35cm}\includegraphics*[height=7.25cm]{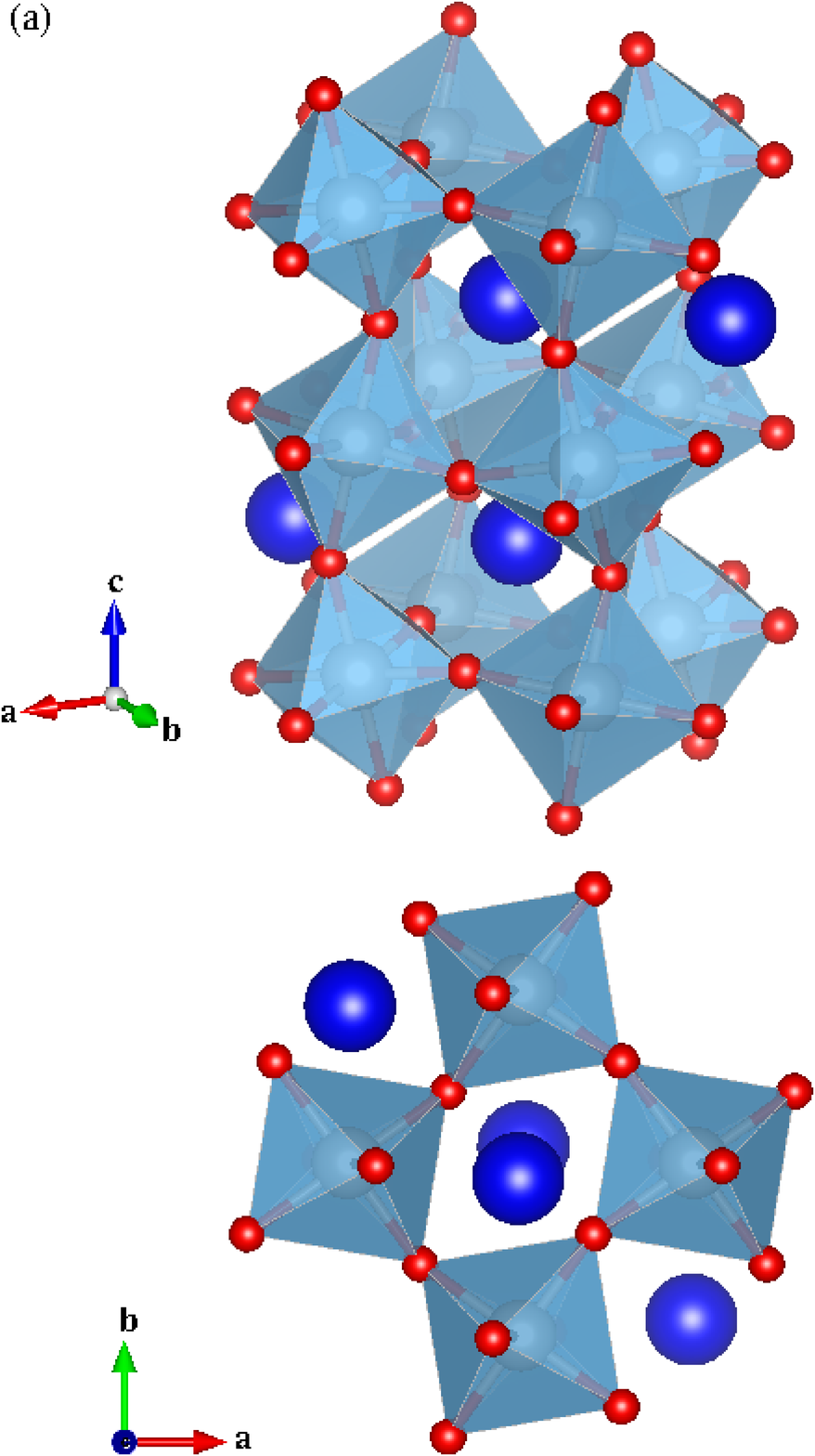}
\hspace*{0.25cm}\includegraphics*[height=8.5cm]{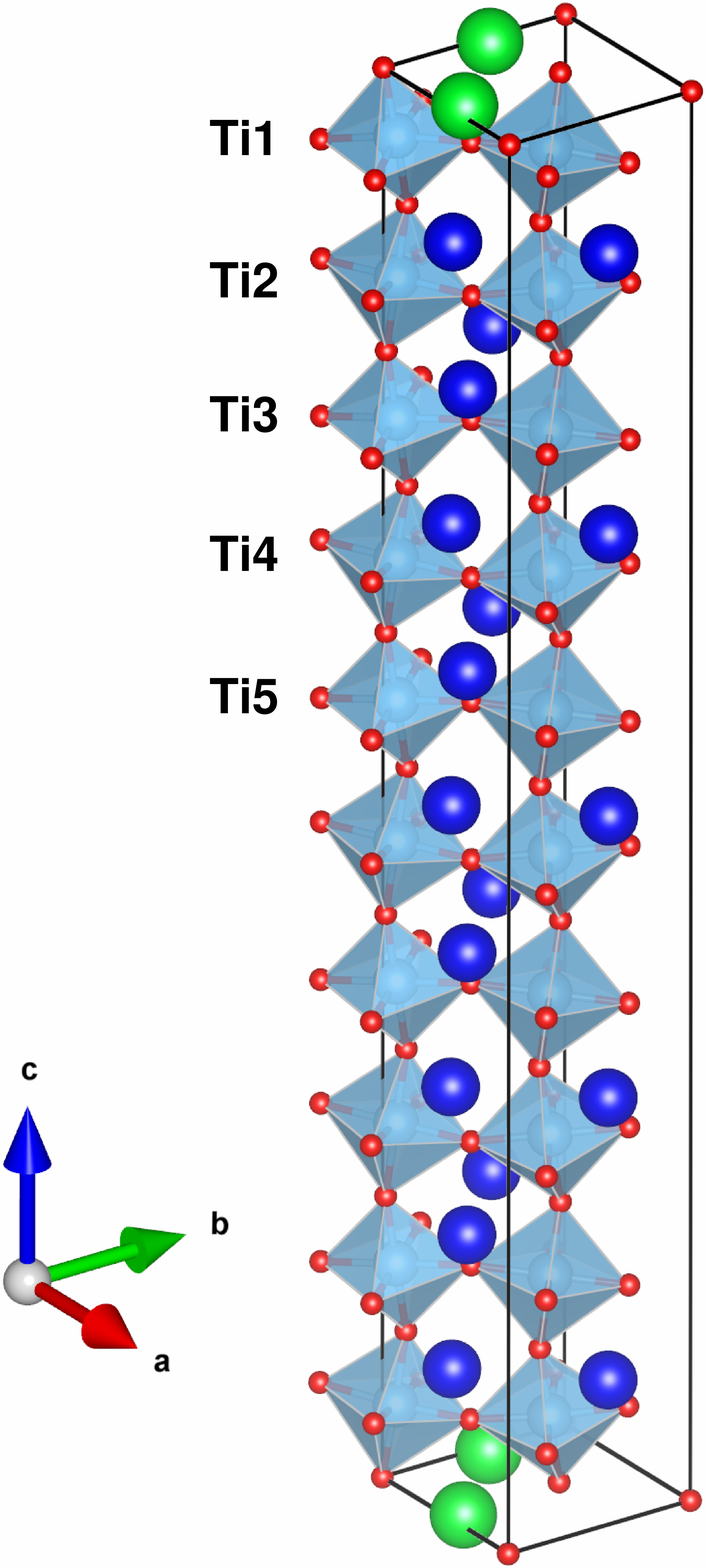}
\hspace*{0.25cm}\includegraphics*[height=8.5cm]{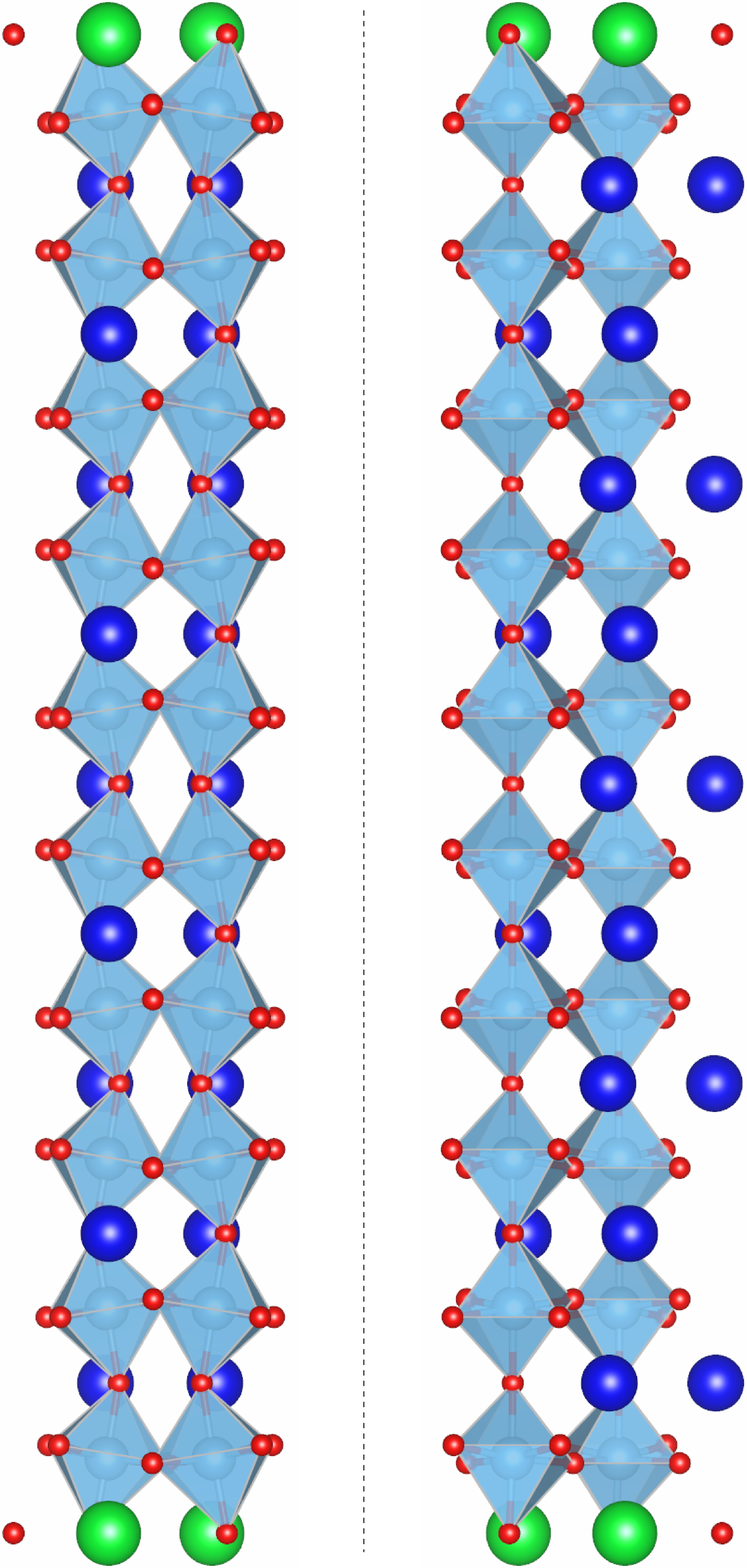}
\end{center}
\vspace*{-0.4cm}
(a)\hspace*{3.75cm}(b)\hspace*{3.75cm}(c)
\caption{ Crystal structures exemplified for the LTO case. 
(a) bulk LTO in perspective view (top) and along the $c$-axis (bottom). (b,c) $\delta$-doped 
SrO/LTO: (b) in perspective view, (c) along $a$-axis (left) and along $b$-axis (right). 
La (blue), Sr (green), Ti (gray), O (red).\label{fig1:structures}}
\end{figure}
%%%%%%%%%%%%%%%%%%%%%%%%%%%%%%%%%%%%%%%%%%%%%%%%%%%%%%%%%%%%%%%%%%%%%%%%%%%%%%%%%%%%%%%%%
In an initial study~\cite{lec13}, the $\delta$-doping of LTO with a single SrO layer was in part
already investigated within DFT+DMFT. But there the inplane lattice constant $a$=$b$ of bulk 
SrTiO$_3$, a site-averaged DC as well as cubic $t_{2g}$ projections, i.e. not adapted
to the local crystal-field eigenbasis, were used. Albeit valuable insight was obtained, the 
correlation physics of such heterostructures is rather sensitive to details and therefore
we here advance in the modelling. Concerning the notorious double-counting term, a
site-dependent DC correction is more tailored to the real-space problem of correlated
sites with different distances to existing interfaces.

\section{Crystal-structure considerations\label{sec:struc}}
Bulk crystal structures of LTO and GTO with $Pbnm$ space group are experimentally well studied and 
we here utilize available x-ray diffraction data~\cite{kom07}
(see Fig.~\ref{fig1:structures}a). Relevant parameters of the distorted pervoskite structure 
with 4 formula units in the primitive cell are tilt and in-plane rotation of the TiO$_6$
octahedra, both with respect to the $c$-axis. The deviations from the perovskite structure
are larger in GTO. A measure thereof is given by the angles between Ti-O1-Ti and Ti-O2-Ti,
where O1 marks the apical oxygen ion and O2 the one in the basal plane of the octahedron. The
latter plane is furthermore distorted differently in both structures. While in LTO that 
distortion is rectangular, in GTO it is of parallelogram shaping~\cite{kom07}.
%%%%%%%%%%%%%%%%%%%%%%%%%%%%%%%%%%%%%%%%%%%%%%%%%%%%%%%%%%%%%%%%%%%%%%%%%%%%%%%%
\begin{figure}[b]
\begin{center}
\includegraphics*[width=11cm]{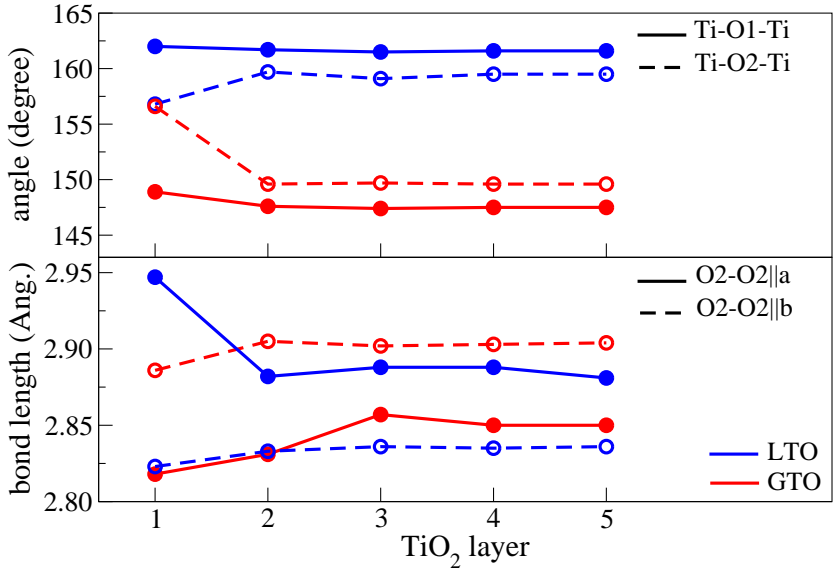}
\end{center}
\caption{Relevant layer-dependent cell-internal paramaters for the $\delta$-doped structures
within LDA. Top: bond angles Ti-O-Ti. Bottom: bond lengths $l$ in the basal plane of the 
TiO$_6$ octahedra. The experimental reference values for the stoichiometric bulk compounds are 
$\angle$Ti-O1-Ti=153.9$^\circ$ (143.9$^\circ$) and 
$\angle$Ti-O2-Ti=153.3$^\circ$ (145.8$^\circ$) for LTO(GTO) as well as $l$(O2-O2$||a$)=$2.94(2.90)$\AA\, and
$l$(O2-O2$||b$)=$2.85(2.91)$\AA\, for LTO(GTO)~\cite{kom07}.\label{fig2:cellintern}}
\end{figure}
%%%%%%%%%%%%%%%%%%%%%%%%%%%%%%%%%%%%%%%%%%%%%%%%%%%%%%%%%%%%%%%%%%%%%%%%%%%%%%%%%%%%%%%%%

To facilitate the $\delta$-doping scenario, a single SrO layer is inserted in the hosting titanates 
within a periodic superlattice architecture in [001] direction, respectively. 
Along the $c$-axis the supercells (see Fig.~\ref{fig1:structures}b) consist of 10 TiO$_2$ and 
9 La(Gd)O host layers inbetween the SrO monolayers. This amounts to a separating distance of about 
4nm between the latter. A $\sqrt{2}$$\times$$\sqrt{2}$ unit is chosen in lateral in-plane 
direction to allow for the TiO$_6$ octahedral tilts/rotations and distortions, i.e.
each TiO$_2$ layer includes two Ti sites. Thus there are 20 Ti ions in these supercells with a 
total number of 100 atoms. For the lattice parameters $a$, $b$ and $c$ the experimental bulk 
LTO/GTO values are used and the atomic positions are obtained from minimizing the atomic 
forces within LDA down to 10 mRyd/a.u. per site. Note that the method of choice for structural 
optimization in low-symmetry correlated materials, i.e. LDA, GGA or static DFT+U~\cite{par12} 
is still open and a matter of debate. 

It is well known that the tilts, rotations and distortions of the TiO$_6$ octahedra in the $R$TiO$_3$
series are crucial for the Mott insulating as well as the magnetic state. Thus the changes thereto
by the SrO doping layer may have strong influence on the resulting electronic structure.
The relaxed structural data is summarized by providing layer-dependent bond angles 
and bond lengths (see Fig.~\ref{fig2:cellintern}). In general,
the LDA bond angles in the $\delta$-doped structures turn out somewhat larger than the experimental
values for the bulk stoichiometric systems. But the hierachy, marking GTO with stronger tilts and
rotations of the TiO$_6$ octahedra, remains intact. For both structures, the differences in tilts
and rotations are strongest close to the SrO doping layer. The octahedral tilt and rotation are 
related to the Ti-O1-Ti, Ti-O2-Ti angles. The former(latter) angle is somewhat enhanced(reduced) 
close the interface in the LTO case. In the GTO case both angles are shifting in the direction of the 
non-distorted value of 180$^{\circ}$ when approaching the interface. Note that these overall smaller 
tilts compared to the bulk cases are in qualitative agreement with recent measurements using scanning 
transmission electron microscopy on SrO quantum wells in GTO~\cite{zha13}. The LDA bond lengths between 
the O2 ions in the octahedral basal plane are in general somewhat lower than in the experimental bulk 
structures. Close to the interface the rectangular distortion is enhanced in the LTO case, while 
for $\delta$-doped GTO some shrinking of the basal plane may be noticed. 

Our structural data is in line with recent GGA+U calculations by Chen {\sl et al.} where a
smaller supercell for $\delta$-doped GTO along the [001] direction was used. On the experimental 
side, Zhang {\sl et al.}~\cite{zha13,oue13,zha14} studied the structural modifications due to doping 
GTO with a SrO monolayer, but using a different interface geometry. There it is defined
by the directions [110] and [001], i.e. the $c$-axis parallel to the interface and $a$,$b$ 
inclined. Whereas in the present work the interface is given by the directions [100] and [010].
The observation of somewhat smaller octahedral tilts right at the interface agrees
between theory and experiment. Still the geometry influence on the electronic structure may be 
significant, e.g. because of orbital-ordered Mott(-like) states with a unique directional 
character.  

\section{Electron states in bulk LaTiO$_3$ and bulk GdTiO$_3$}
%%%%%%%%%%%%%%%%%%%%%%%%%%%%%%%%%%%%%%%%%%%%%%%%%%%%%%%%%%%%%%%%%%%%%%%%%%%%%%%%%%%%%%%%%
\begin{figure}[b]
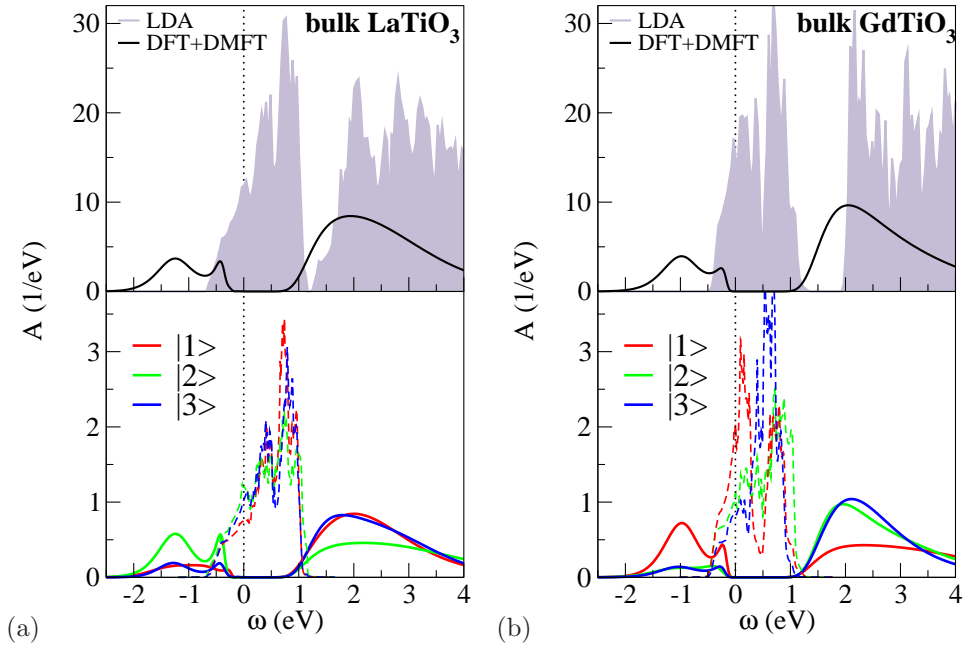

\begin{center}
(a)\hspace*{-0.25cm}\includegraphics*[width=6cm]{fullspec-lto.eps}\hspace*{0.2cm}
(b)\hspace*{-0.25cm}\includegraphics*[width=6cm]{fullspec-gto.eps}
\end{center}
\caption{ Spectral-function comparison between LDA and DFT+DMFT ($T$=290K)
for the bulk case of (a) LTO and (b) GTO. Top: total spectrum, bottom: local orbital-resolved 
spectrum with dashed lines for the LDA result.\label{fig2:bulk-data}}
\end{figure}
%%%%%%%%%%%%%%%%%%%%%%%%%%%%%%%%%%%%%%%%%%%%%%%%%%%%%%%%%%%%%%%%%%%%%%%%%%%%%%%%%%%%%%%%%
This section summarizes the electronic structure of the bulk compounds to set the stage for
comparison with the $\delta$-doped cases. The La- and Gd-titanate are well-defined metals in 
the conventional Kohn-Sham representation of DFT. Both display 
an $t_{2g}$-like low-energy manifold close to the Fermi level $\varepsilon_{\rm F}$ in LDA 
(see Fig.~\ref{fig2:bulk-data}) of bandwidths $W_{\rm LTO}$=1.9eV  and $W_{\rm GTO}$=1.8eV. 
The GTO density of states (DOS) at $\varepsilon_{\rm F}$ is larger than the LTO one and the 
$t_{2g}$-$e_g$ gap in the unoccupied energy region is also increased for the former. Therefrom
the Gd compound is somewhat stronger susceptible to correlation effects than the La one.
%%%%%%%%%%%%%%%%%%%%%%%%%%%%%%%%%%%%%%%%%%%%%%%%%%%%%%%%%%%%%%%%%%%%%%%%%%%%%%%%%%%%%%%%%
\begin{figure}[t]
\begin{center}
\includegraphics*[width=12cm]{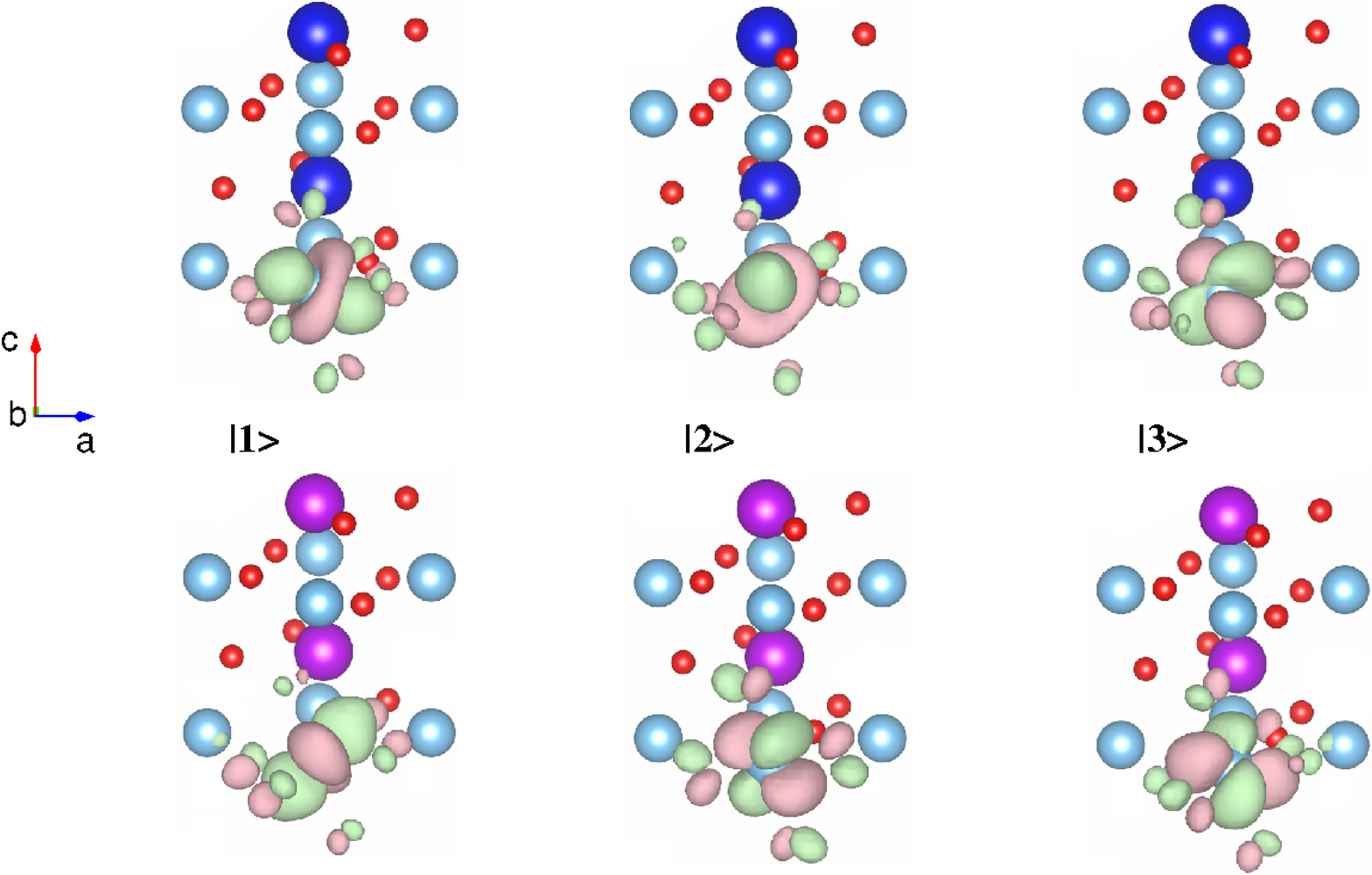}
\end{center}
\caption{ Crystal-field orbitals from the projection formalism for the bulk cases
of LTO (top) and GTO (bottom). La (blue), Gd( violet), Ti (gray), O (red).\label{fig3:orbitals}}
\end{figure}
%%%%%%%%%%%%%%%%%%%%%%%%%%%%%%%%%%%%%%%%%%%%%%%%%%%%%%%%%%%%%%%%%%%%%%%%%%%%%%%%%%%%%%%%%

The LTO local crystal-field basis from the projected local orbitals reads (note that phases
differ on the various Ti sites)
\begin{equation}
\left(\begin{array}{r} \ket{1} \\ \ket{2} \\ \ket{3} \end{array} \right)=
\left(\begin{array}{rrr}
 0.710 &  0.274 & -0.648 \\
 0.262 &  0.752 &  0.605 \\
 0.653 & -0.600 &  0.462
\end{array} \right)
\left(\begin{array}{r} xz \\ yz \\ xy \end{array} \right),\label{eq:ltotraf}
\end{equation}
with level energies $\varepsilon_{\rm cf}^{(1)}$=483meV, $\varepsilon_{\rm cf}^{(2)}$=379meV and 
$\varepsilon_{\rm cf}^{(3)}$=491meV. 
Lowest in energy is state $\ket{2}$ with dominant $yz$ and weakest $xz$ contribution, 
in qualitative agreement with the previous study~\cite{pav05} based on linear/$N$th-order 
muffin-tin orbital methods in the atomic-sphere approximation. Crystal-field splittings 
$\Delta_{2,3}$=$\varepsilon_{\rm cf}^{(2,3)}-\varepsilon_{\rm cf}^{(1)}$ of the order of 100meV 
turn out to be somewhat smaller in the present work. The local DOS of these states is plotted 
in Fig.~\ref{fig2:bulk-data} and shows no very significant orbital discrimination in its shape. 
No sizeable orbital polarization is seen on the LDA level, the single Ti$(3d)$ electron is shared 
among the adapted states.

For GTO the corresponding local crystal-field basis is given by
\begin{equation}
\left(\begin{array}{r} \ket{1} \\ \ket{2} \\ \ket{3} \end{array} \right)=
\left(\begin{array}{rrr}
0.639  &  0.449 &  0.625 \\
0.707  & -0.664 & -0.245 \\
-0.305 & -0.598 &  0.741
\end{array} \right)
\left(\begin{array}{r} xz \\ yz \\ xy \end{array} \right),\label{eq:gtotraf}
\end{equation}
with level energies $\varepsilon_{\rm cf}^{(1)}$=319meV, $\varepsilon_{\rm cf}^{(2)}$=483meV and 
$\varepsilon_{\rm cf}^{(3)}$=456meV.
State $\ket{1}$ with lowest energy is here a more balanced combination of the original $t_{2g}$
orbitals, with weakest contribution from $yz$. The crystal-field splittings are similar 
to the LTO case, with a small increase in numbers. Stronger orbital discrimination is visible on 
the LDA level, i.e. the local DOS of the $\ket{1}$ state marks not only a smaller bandwidth but has 
a significant larger value at the Fermi level. This leads to some orbital polarization towards 
$\ket{1}$ in the single-electron share.

Figure~\ref{fig3:orbitals} shows the obtained effective $t_{2g}$ orbitals for both titanates.
As within other downfolding schemes applied to transition-metal oxides~\cite{pav05,lec06}, the 
Wannier-like functions exhibit significant weight on the nearby oxygen ions, as expected via 
the projection from low-energy KS states. Note that though the projection orbitals are atomic-like
functions~\cite{ama08}, the resulting correlated subspace is composed of Wannier-like functions 
due to the projection onto selected KS states. Concerning the lowest-energy crystal-field 
orbitals, the $\ket{2}$ state of LTO is mainly oriented along the $b$-axis, while the $\ket{1}$
state of GTO is aligned along the $a$-axis. Furthermore the GTO $\ket{3}$ orbital has a clearer
in-plane $xy$ character, whereas a clear resemblance thereof is lacking in LTO.

The electronic structure of these compounds is modified substantially within DFT+DMFT. For both
materials the experimental room-temperature Mott-insulating state is confirmed 
(see Fig.~\ref{fig2:bulk-data}). Lower Hubbard bands are located at $\sim$1.2eV for LTO and at 
$\sim$1eV for GTO. In our present scheme the charge gaps amount to 
$\Delta^{\rm g}_{\rm LTO}$$\sim$0.5eV and $\Delta^{\rm g}_{\rm GTO}$$\sim$0.8eV. While the latter 
value is in good agreement with experimental data of about 0.75eV~\cite{cra92}, our result 
for the LTO value is larger than $\Delta^{\rm g}_{\rm LTO}$$\sim$0.2-0.3eV from 
experiment~\cite{ari93,oki95}. 
A smaller theoretical charge gap was obtained in previous non-charge-self-consistent 
studies~\cite{pav04} with similar correlated subspace and same interaction parameters, but at 
much higher temperature $T$=1200K. Recent DFT+DMFT with larger correlated subspace and including 
the unoccupied $4f$ state reaches a smaller value also close to room temperature~\cite{hau14}. 
Substantial orbital polarization takes place in the Mott state, with dominant filling of  
the effective $\ket{2}$ ($\ket{1}$) orbital in LTO (GTO). The occupation numbers for
$(\ket{1},\ket{2},\ket{3})$ read (0.22, 0.59, 0.19) for LTO and (0.66, 0.18, 0.16) for GTO. Hence
with correlations, close to 2/3 of the single $t_{2g}$ electron resides in lowest crystal-field 
level.

\section{Electron states in  $\delta$-doped compounds without 
broken spin/charge symmetry\label{sec:dftdmft-delta}}
Realistic correlation effects from $\delta$-doping LTO and GTO are approached in a 
multi-site DFT+DMFT scheme. The 5 inequivalent Ti ions in the supercell structures allow for
corresponding layer-dependent effective-$t_{2g}$ DMFT self-energies 
$\mathbf\Sigma$1$-$5 beyond static considerations. In this section we first concentrate on the 
electronic structure without possible breaking of spin and charge symmetries, i.e. paramagnetic 
states as well as charge-balanced solutions. To this, no intra-layer discrimination of the both Ti
ions per layer is performed, respectively. Symmetry-broken solutions for $\delta$-doped GTO 
will be discussed in section~\ref{sec:delta-order}.

Table~\ref{tab:projections} provides the list of matrices, based on LDA calculations, transforming 
the original $(xz, yz, xy)$ orbitals into the, now layer-dependent, crystal-field bases 
$(\ket{1}, \ket{2}, \ket{3})$. 
The variations in the coefficients are largest between the groups (Ti1,Ti2) and (Ti3,Ti4,Ti5), but the
classification of the effective orbitals remains stable across the different TiO$_2$ layers.
From the onsite level energies $\varepsilon_{\rm cf}$ of the effective crystal-field states it is seen
that again the $\ket{2}$($\ket{1}$) state is lowest on the Ti3-5 sites in the doped LTO(GTO)
case, as found for the bulk Ti site. The level spacing on those three sites is also nearly
identical to the bulk for GTO, whereas for the LTO case the energy difference is up to 60\%
larger. On Ti2 the level energies come already substantially closer compared to Ti3-5.
Qualitative energetic changes occur for the TiO$_2$ layer closest to the SrO doping
layer: the Ti1 sites has state $\ket{3}$, with dominant $xy$ contribution, lowest for
both titanates types.
%%%%%%%%%%%%%%%%%%%%%%%%%%%%%%%%%%%%%%%%%%%%%%%%%%%%%%%%%%%%%%%%%%%%%%%%%%%%%%%%
\begin{table}[b]
%\footnotesize
%\setlength{\extrarowheight}{1.5pt}
\begin{tabular}{lr|r r r cc|r r r cc}
        &           &  & $\delta$-LTO   &  & & & & $\delta$-GTO   & & \\ 
    &    & $xz$      & $yz$  &  $xy$  & & $\varepsilon_{\rm cf}$  & $xz$ & $yz$  &  $xy$  & & $\varepsilon_{\rm cf}$ \\ \hline
    & $\ket{1}$  & \squ -0.930&  0.212& -0.300\squ   & &656 & \squ -0.735&  0.227& -0.638\squ&& 668  \\
Ti1 & $\ket{2}$  & \squ -0.354& -0.739&  0.573\squ   & &663 & \squ -0.328& -0.944&  0.041\squ&& 730  \\
    & $\ket{3}$  & \squ  0.100& -0.639& -0.763\squ   & &597 & \squ  0.593& -0.240& -0.769\squ&& 577  \\ \hline
    & $\ket{1}$  & \squ  0.943& 0.322 & 0.086\squ    & &569  & \squ  0.746& -0.268&  0.609\squ&& 409  \\ 
Ti2 & $\ket{2}$  & \squ -0.176& 0.699 &-0.693\squ    & &428  & \squ  0.362&  0.932& -0.032\squ&& 534  \\
    & $\ket{3}$  & \squ -0.284& 0.638 & 0.716\squ    & &587  & \squ -0.559&  0.245&  0.792\squ&& 508   \\ \hline
    & $\ket{1}$  & \squ -0.984&  0.168&  0.056\squ   & &545  & \squ 0.718 & 0.321 & 0.617\squ && 387  \\ 
Ti3 & $\ket{2}$  & \squ -0.165& -0.754& -0.636\squ   & &387  & \squ 0.324 & -0.939&  0.111\squ&& 524   \\
    & $\ket{3}$  & \squ -0.065& -0.635&  0.770\squ   & &577  & \squ 0.616 & 0.120 & -0.779\squ&& 502   \\ \hline
    & $\ket{1}$  & \squ -0.982&  0.190& 0.002\squ    & &544  & \squ -0.715& -0.331&  0.616\squ&& 382  \\ 
Ti4 & $\ket{2}$  & \squ -0.146& -0.759& 0.634\squ    & &390  & \squ -0.289&  0.942&  0.170\squ&& 519  \\
    & $\ket{3}$  & \squ  0.122&  0.622& 0.773\squ    & &578  & \squ -0.637& -0.057& -0.769\squ&& 498  \\ \hline
    & $\ket{1}$  & \squ -0.978&  0.210&  0.012\squ   & &547  & \squ 0.721 & 0.325 & 0.612\squ && 384  \\ 
Ti5 & $\ket{2}$  & \squ -0.171& -0.758& -0.629\squ   & &395  & \squ 0.305 & -0.942&  0.140\squ&& 521    \\
    & $\ket{3}$  & \squ -0.123& -0.617&  0.777\squ   & &575  & \squ 0.622 & 0.086 & -0.779\squ&& 499   \\ 
\end{tabular}
%\end{ruledtabular}
\caption{Layer-dependent Ti crystal-field bases within the projected local orbitals defining 
the correlated subspace for the $\delta$-doped compounds. The given 3$\times$3 matrices 
transform the $(xz, yz, xy)$ orbitals to the tailored $(\ket{1}, \ket{2}, \ket{3})$ orbitals in 
the same manner as given in  eqs. (\ref{eq:ltotraf},\ref{eq:gtotraf}) for the respective bulk 
case. The quantities $\varepsilon_{\rm cf}$ denote the respective onsite level energies (in meV) for the
resulting tailored orbitals.\label{tab:projections}}
\end{table}
%%%%%%%%%%%%%%%%%%%%%%%%%%%%%%%%%%%%%%%%%%%%%%%%%%%%%%%%%%%%%%%%%%%%%%%%%%%%%%%%%%%%%%%%%
\begin{figure}[t]
\begin{center}
(a)\includegraphics*[width=10cm]{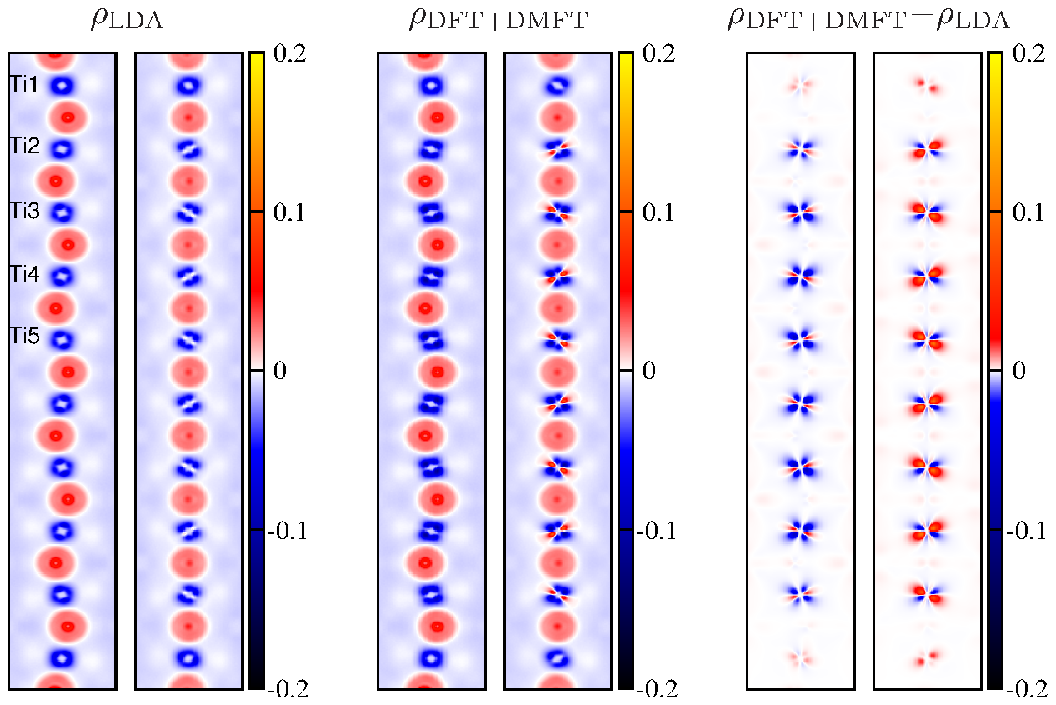}\\[0.2cm]
(b)\includegraphics*[width=10cm]{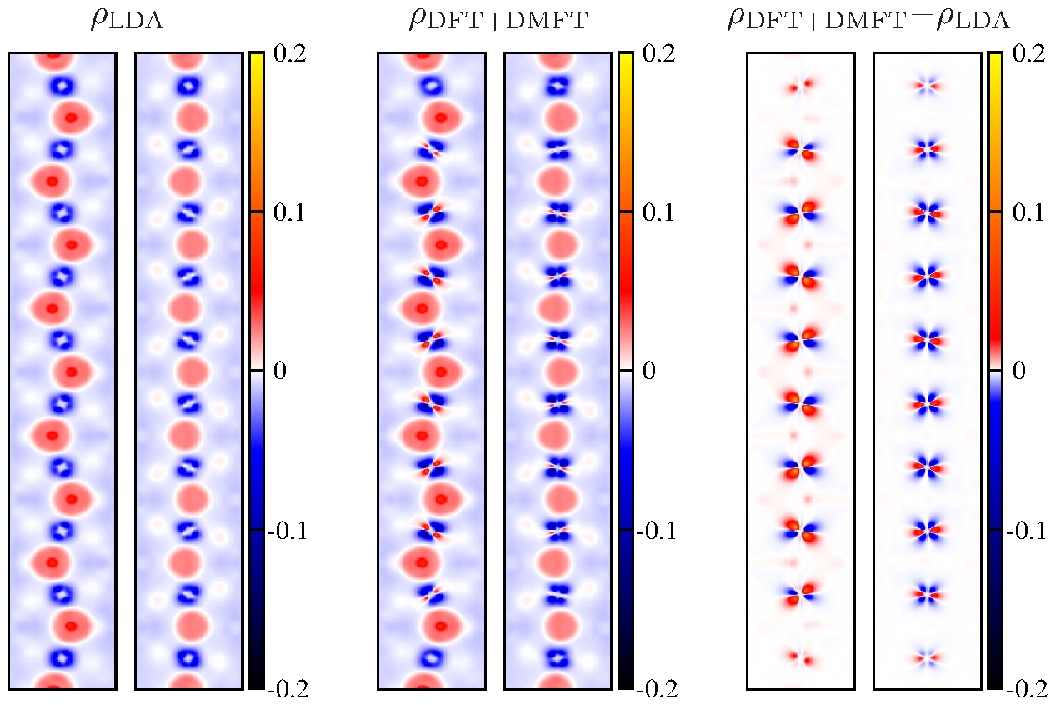}
\end{center}
\caption{ Bond charge densities $\rho$=$\rho_{\rm total}$$-$$\rho_{\rm atomic}$ for
$\delta$-doped (a) LTO and (b) GTO. For each protocol, densities within (left) the $ac$-plane and
(right) the $bc$-plane are given. The DFT+DMFT data is retrieved from calculations at 
$T$=290K.\label{fig4:charges}}
\end{figure}
%%%%%%%%%%%%%%%%%%%%%%%%%%%%%%%%%%%%%%%%%%%%%%%%%%%%%%%%%%%%%%%%%%%%%%%%%%%%%%%%%%%%%%%%%

%%%%%%%%%%%%%%%%%%%%%%%%%%%%%%%%%%%%%%%%%%%%%%%%%%%%%%%%%%%%%%%%%%%%%%%%%%%%%%%%%%%%%%%%%
\begin{figure}[b]
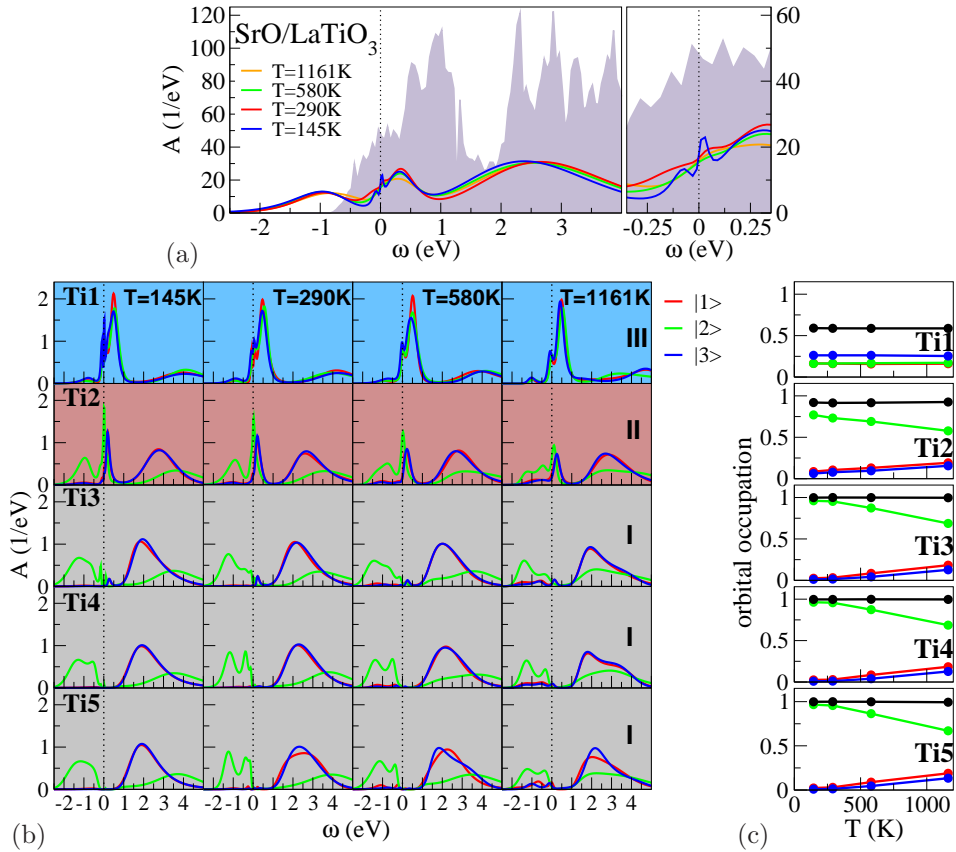

\begin{center}
(a)\hspace*{-0.5cm}\includegraphics*[width=8.5cm]{ldadmft-dos.lto.eps}\\[0.1cm]
(b)\hspace*{-0.5cm}\includegraphics*[height=7.5cm]{lto-sto.locti.eps}
(c)\hspace*{-0.5cm}\includegraphics*[height=7.5cm]{occ.lto.eps}
\end{center}
\caption{ $k$-integrated spectral function $A(\omega)$ for $\delta$-doped LTO
from DFT+DMFT. (a) Total function compared to LDA (bluegrey), where $A(\omega)$ 
builds on the 60 electron states from the bottom of the low-energy $t_{2g}$ manifold. Left:
blow-up around $\varepsilon_{\rm F}$. 
(b) Local Ti-resolved functions, with three electronic regions: orbital-ordered Mott insulating 
(grey bg), orbital-polarized doped (lightred bg) and orbital-selective doped (lightblue bg). 
(c) Ti-resolved orbital fillings, black datapoints/line marks the respectve total 
$n$.\label{fig:sro/lto}}
\end{figure}
%%%%%%%%%%%%%%%%%%%%%%%%%%%%%%%%%%%%%%%%%%%%%%%%%%%%%%%%%%%%%%%%%%%%%%%%%%%%%%%%%%%%%%%%%
\begin{figure}[t]
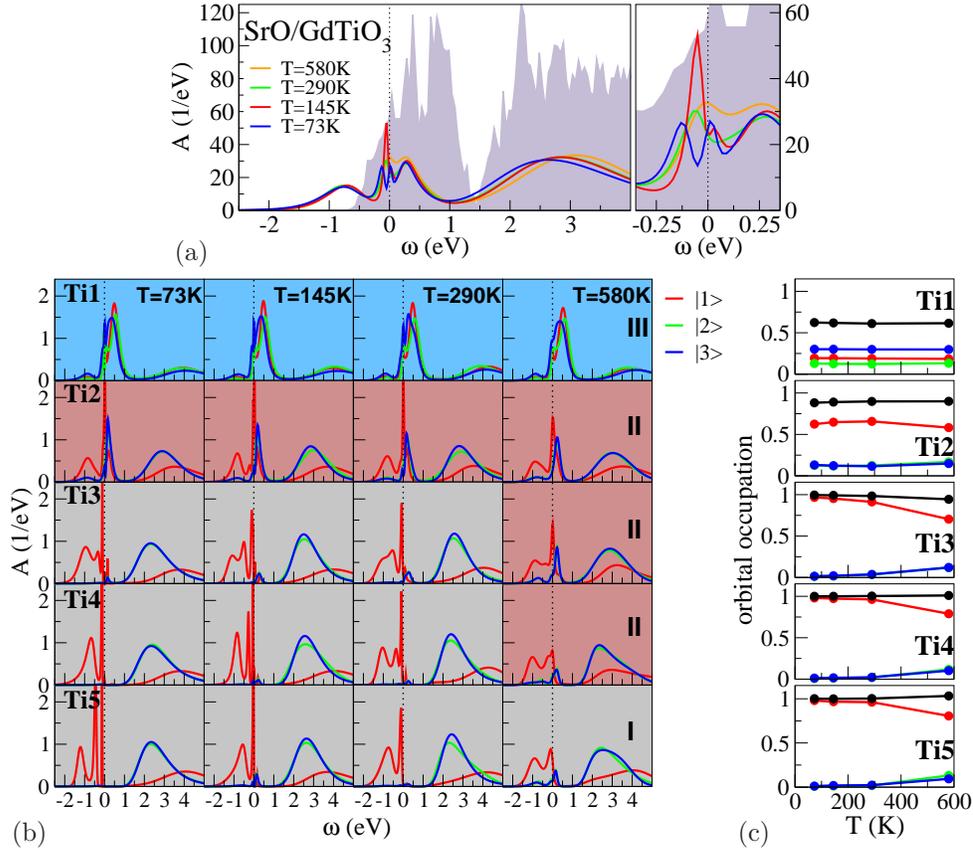

\begin{center}
(a)\hspace*{-0.5cm}\includegraphics*[width=8.5cm]{ldadmft-dos.gto.eps}\\[0.1cm]
(b)\hspace*{-0.5cm}\includegraphics*[height=7.5cm]{gto-sto.locti.eps}
(c)\hspace*{-0.5cm}\includegraphics*[height=7.5cm]{occ.gto.eps}
\end{center}
\caption{ $k$-integrated spectral function $A(\omega)$ for $\delta$-doped GTO
from DFT+DMFT. The subfigures (a-c) are described as in Fig.~\ref{fig:sro/lto}.\label{fig:sro/gto}}
\end{figure}
%%%%%%%%%%%%%%%%%%%%%%%%%%%%%%%%%%%%%%%%%%%%%%%%%%%%%%%%%%%%%%%%%%%%%%%%%%%%%%%%%%%%%%%%%
The single SrO layer induces hole doping in the $3d^1$ titanates, trying to shift the nominal 
Ti$^{3+}$ state towards Ti$^{4+}$ as found in bulk SrTiO$_3$. Since the complete unit cell hosts 
ten TiO$_2$ layers the doping amounts to 0.1 electrons per Ti ion. Figure~\ref{fig4:charges}
displays the resulting bond charge density (BCD) within LDA and DFT+DMFT. These densities
are evaluated as differences between the obtained total charge density and the original overlayed
atomic charge densities, thus reveal the key effects due to crystallization. One main BCD feature 
for the considered transition-metal-oxide compounds is the ionic charge transfer from titanium to
oxygen. Beyond that the LDA BCD does not show much more deeper characteristics. The BCD from
DFT+DMFT on the other hand clearly reveals the orbital polarization in the different 
crystallographic directions for LTO and GTO far from the SrO layer. A difference plot
visualizes this polarization readily as local correlation effects, i.e. onsite $3d$ charge
transfers. Especially in the GTO case, further correlation-induced charge transfer to 
oxygen ions occurs. Close to the SrO layer the orbital polarization is weak, mainly a minor
additonal charge transfer into that region from the interior within DFT+DMFT compared to LDA
becomes visible.

Further details on the $\delta$-doped LTO correlated electronic structure are revealed from the $k$-summed
one-particle spectral function $A(\omega)$=$\sum_{\bf k}A({\bf k},\omega)$. The studied
$T$ range starts above 1000K and ends at 145K, close to the bulk Ne{\'e}l temperature.
The total spectrum has metallic character for all investigated temperatures, but exhibits 
strong transfer of spectral weight to Hubbard peaks (see Fig.~\ref{fig:sro/lto}a). 
Compared to the bulk case, the 
centre of the lower Hubbard peak is shifted upwards to $\sim$0.9-1eV. The quasiparticle (QP) 
peaks have significant $T$ dependence, displaying bad-metal behavior at eleveated temperatures. 
Only well below room temperature a clear resonance occurs at the Fermi level. A
layer-dependent multi-orbital Mott transition is infered from the local 
Ti-resolved $A(\omega)$. Its structure renders it possible to separate the $\delta$-doped 
electronic structure into three distinct regimes (see Fig.~\ref{fig:sro/lto}b): Far from the 
doping layer the material is in a {\sl orbital-ordered Mott state} (I), with insulating layers derived 
from Ti3-5. Notably these Mott layers are still different from the bulk-LTO state. The orbital 
polarization towards (layer-dependent) $\ket{2}$ is nearly complete already close to room 
temperature (compare also Fig.~\ref{fig:sro/lto}c), while in bulk LTO only about 2/3 of the single 
electron resides in state $\ket{2}$. This is in line with the finding of stronger energy separation
of the onsite crystal-field levels in $\delta$-LTO (cf. Tab.~\ref{tab:projections}).
In addition, the local charge gap is somewhat reduced compared 
to the bulk value. For higher $T$ the gaps are partially filled by incoherent excitations. The 
second regime (II) is build from the next-nearest TiO$_2$ layer from the interface, an 
{\sl orbital-polarized doped-Mott} layer with QPs at low energy. While the occupied part, including 
Hubbard and QP excitation, is dominated by $\ket{2}$, minor QP weight is shifted towards $\ket{1}$, 
$\ket{3}$. Finally the third regime (III) is formed by the nearest TiO$_2$ layer right below SrO,
an {\sl orbital-selective} metallic film with comparatively more orbital-balanced character but 
clearly favoring the $xy$-dominated $\ket{3}$ state in the occupied part. The coherence of the QPs 
therein increases significantly at low $T$, while the orbital fillings are rather temperature 
independent. About 0.6 electrons are located at Ti1. The remaining holes are mostly at Ti2, but 
the hole doping is not fully perfect, i.e. some holes seem to remain within the SrO layer.
%%%%%%%%%%%%%%%%%%%%%%%%%%%%%%%%%%%%%%%%%%%%%%%%%%%%%%%%%%%%%%%%%%%%%%%%%%%%%%%%%%%%%%%%%
\begin{figure}[t]
\begin{center}
\includegraphics*[width=8cm]{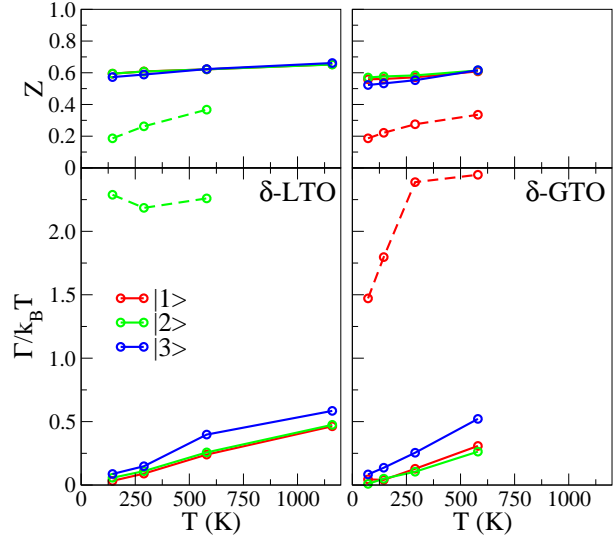}\\[0.1cm]
\end{center}
\caption{ Orbital-resolved quasiparticle weight $Z$ and dimensionless scattering rate 
$\Gamma/k_{\rm B}T$ for the $\delta$-doped compounds. Full lines correspond to the states right 
at the SrO layer and the dashed line to the dominant doped Mott state in the second-distant 
layer.\label{fig:sigdat}}
\end{figure}
%%%%%%%%%%%%%%%%%%%%%%%%%%%%%%%%%%%%%%%%%%%%%%%%%%%%%%%%%%%%%%%%%%%%%%%%%%%%%%%%%%%%%%%%%

Accordingly, the spectral data for $\delta$-doped GTO is exhibited in Fig.~\ref{fig:sro/gto}. Since
the Curie temperature for ferromagnetic ordering of bulk GTO is well below 100K, we shifted the
temperature range and set $T$=73K as the lowest value. In comparison to $\delta$-LTO, the QP 
coherence close to the Fermi level is enlarged, but shows a surprising decline below 100K. 
In general, the temperature dependence is more delicate in $\delta$-doped gadolinium titanate, as 
can be also seen in the local Ti-resolved $A(\omega)$. The three encountered electronic regimes are 
also present here, now with the corresponding state $\ket{1}$ as the Mott-dominating one. But for 
Ti3 and Ti4 there is in addition an $T$-dependent transition between regimes I and II. Furthermore 
the gapped QP of $\ket{3}$ in the occupied region has a much stronger signature 
and resides just below the Fermi level. The orbital-occupation character far away from the interface 
resembles the LTO case, here strongly polarized in the $\ket{1}$ state with clear $3d^1$ total
filling. Again the strength of orbital polarization is significantly increased compared to the GTO
bulk case. Note that here, since the LDA onsite level differences are similar to the bulk, the 
DMFT self-energy is nearly exclusively responsible for that increase. In general the 
correlation-induced local level shift renormalizing the original crystal-field splitting is given by 
$\Delta_{\rm c}$=$\varepsilon_{\rm CSC}$$+$$\mbox{Re}\,\Sigma(i0^+)$$-$$\Delta_{\rm DC}$$-$$\varepsilon_{\rm cf}$.
Here $\varepsilon_{\rm CSC}$ is the local level energy from the KS-like part in
DFT+DMFT, $\Delta_{\rm DC}$ is the orbital-independent but site-dependent shift from
the fully-localized double counting and $\varepsilon_{\rm cf}$ the original crystal-field
level energy from LDA. In the close-to-interface regimes II and III there are slight differences, namely 
marginal depletion of the doped-Mott state at lower $T$ in II and somewhat stronger orbital 
discrimination in III. The total filling at the interface is similar but marginally increased, i.e.
Ti$^{0.62+}$ compared to Ti$^{0.59+}$ in $\delta$-doped LTO.

In order to provide insight in the correlation strength within the two metallic layers, 
Fig.~\ref{fig:sigdat} finally shows the QP weight $Z$ and the dimensionless electron-electron 
scattering $\Gamma/k_{\rm B}T$ for both doped titanates. The data is retrieved from the respective 
layer-dependent self-energies via
$Z$=$(1-\frac{\partial\mbox{\scriptsize Im}\,\Sigma(i\omega)}{\partial\omega}|_{\omega\rightarrow 0^+})^{-1}$=$\frac{m_{\rm LDA}}{m^*}$ and $\Gamma$=$-k_{\rm B}T\,Z\,\mbox{Im}\,\Sigma(i0^{+})$.
Within the LTO interface layer the QP renormalization is weakly temperature dependent
and with $Z$$\sim$0.6 still moderate. Also for the scattering rate the relation
$\Gamma/k_{\rm B}T$$<$1 holds up to high $T$, marking the regime III as coherent. Albeit
true Fermi-liquid behavior, i.e. $\Gamma/k_{\rm B}T$ linear in $T$, is only expected for very
small temperatures well below the present range, the close-to-linear characteristics is notable.
On the other hand the doped-Mott state in the next-distant layer has a stronger $T$-dependent
QP weight, getting as low as $Z$$\sim$0.2 at $T$=145K. Electron-electron scattering is here
severe and strongly incoherent even at low $T$. Similar observations hold for $\delta$-doped GTO,
with overall somewhat increased correlation strength. Interestingly, the incoherent regime II
seems on the way of restoring coherency at smaller temperatures, but $\Gamma/k_{\rm B}T$ for
the similar $\ket{1}$ character right at the interface deviates from quasi-linear behavior at very
low $T$.

\section{Electron states in  $\delta$-doped GdTiO$_3$ with broken spin symmetry\label{sec:delta-order}}
Finally we discuss possible ordering instabilities in the particle-hole channel. Both bulk 
materials LaTiO$_3$ and GdTiO$_3$ exhibit magnetic ordering at low temperatures, but because of
the complexity of studying electronic ordering in the doped compounds we here restrict
the investigation to only one system, namely the GTO case.
In the bulk, the compound becomes ferromagnetic (FM) below 36K and the strongly enhanced spectral 
weight close to the Fermi level in the $\delta$-doped case might be indicative of a Stoner-like 
magnetic instability. Furthermore, when doped with a SrO monolayer, roughly half of an electron is 
located at Ti ions right at the doping layer which raises the question about in-plane
Ti$^{3+}$/Ti$^{4+}$ charge ordering~\cite{pen07}. 
Concerning charge disproportionation or other possible intra-layer instabilities, we performed 
post-processing (or one-shot) DMFT based on the given projected multi-orbital onsite Hubbard 
Hamiltonian, now with intra-layer discrimination of the both Ti ions, respectively. However no
checkerboard-type or other insulating instability was detected. Nonlocal Coulomb interactions may 
be relevant to trigger such orderings. A charge self-consistent treatment is not performed since it 
asks for a symmetry lowering also in the DFT part, which renders the already heavy calculations 
very expensive. Therefore we still cannot fully exclude the possibility of an insulating solution.
In the following we concentrate on 
broken spin-symmetry and shed light on a possible (in-plane) FM $\delta$-doped state
from a [100]-[010] interface architecture. Experimentally, ferromagnetism in SrTiO$_3$/GdTiO$_3$ 
quantum wells within a [110]-[001] interfacing has indeed been reported by Jackson and 
Stemmer~\cite{jac13}.
%%%%%%%%%%%%%%%%%%%%%%%%%%%%%%%%%%%%%%%%%%%%%%%%%%%%%%%%%%%%%%%%%%%%%%%%%%%%%%%%%%%%%%%%%
\begin{figure}[b]
\begin{center}
(a)\hspace*{-0.5cm}\includegraphics*[width=8.5cm]{sp-ldadmft-dos.gto.eps}\\[0.1cm]
(b)\hspace*{-0.05cm}\includegraphics*[height=7.75cm]{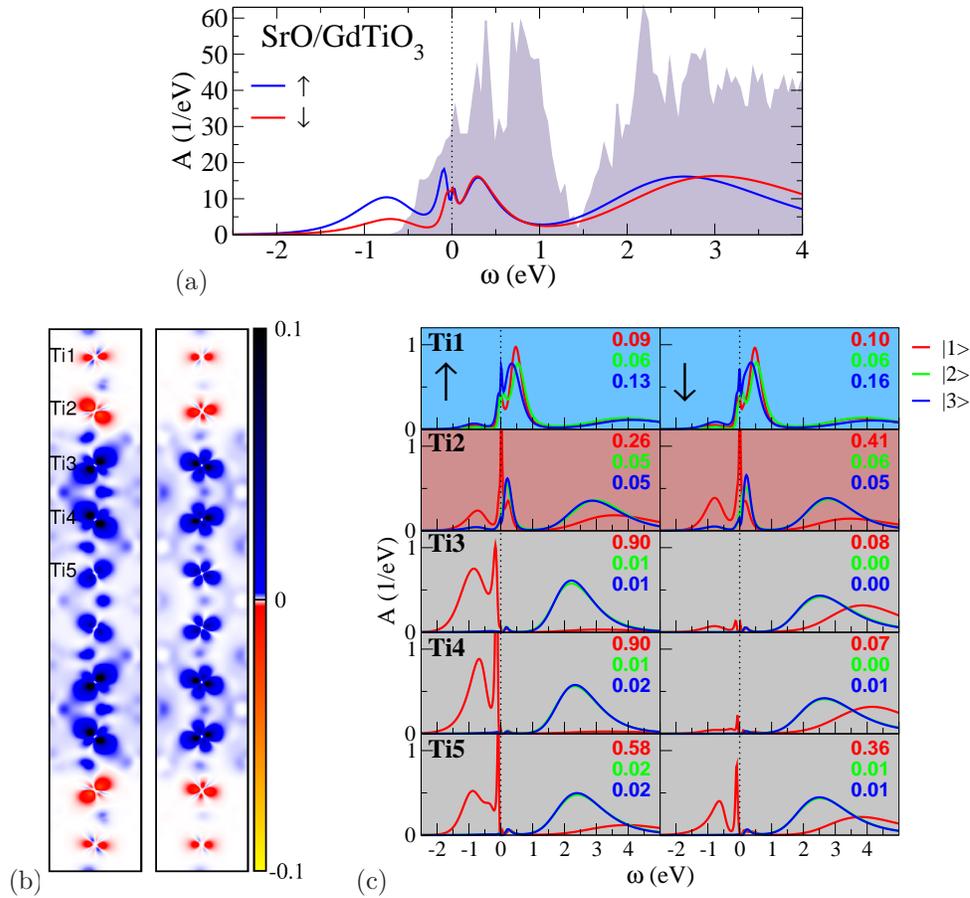}\hspace*{0.5cm}
(c)\hspace*{-0.15cm}\includegraphics*[height=7.5cm]{sp-gto-sto.locti.eps}
\end{center}
\caption{ Ferrimagnetic $\delta$-doped GTO within DFT+DMFT. (a) total spin-resolved
spectral function  compared to LDA (bluegrey). (b) Spin-resolved charge difference 
$\rho_{\uparrow}$$-$$\rho_{\downarrow}$ within the $ac$-plane (left) and the $bc$-plane (right). 
(c) Ti-resolved local spectral function with numbers providing the filling in each orbital-spin
channel.
\label{fig:fm-sro/gto}}
\end{figure}
%%%%%%%%%%%%%%%%%%%%%%%%%%%%%%%%%%%%%%%%%%%%%%%%%%%%%%%%%%%%%%%%%%%%%%%%%%%%%%%%%%%%%%%%%

In principle there are two magnetically active shells in GTO, namely Gd($4f$) and Ti($3d$).
Experimental work suggests that in the bulk an antiferromagnetic (AFM) coupling between Gd and 
Ti sets in at low $T$, giving rise to ferrimagnetism~\cite{tur80,zho05}. 
But since the Gd-O-Gd and Gd-O-Ti exchange is 
believed to be weaker than the (ferromagnetic) Ti-O-Ti one~\cite{tur80}, we here approximate the 
problem by assuming the Gd sites paramagnetic without taking part in the formation of magnetic 
order. This is reasonable since the temperature $T$=145K is also set way above the bulk 
$T_{\rm C}$=36K.

Post-processing (or one-shot) DMFT starting from the charge self-consistent paramagnetic DFT+DMFT
solution does not lead to FM order. Using charge-only self-consistency, i.e. neglecting spin
polarization in the KS-DFT part, also results in vanishing ordered moments. Only the complete
spin-resolved charge self-consistent treatment sustains magnetic order when starting from 
FM initialization at $T$=145K. Note that the latter protocol may overestimate magnetic-ordering 
tendencies, because once an exchange-splitting is introduced in KS-DFT via the DMFT Ti1-5 
self-energies it may not easily suppressed within the formalism. On the other hand taking into 
account spin polarization in the KS part may be vital. Moments on the ligands and corresponding 
exchange is important for the overall lattice magnetic ordering.

A total moment of 6.8$\mu_{\rm B}$ is obtained for the whole 100-atom supercell with the 20 Ti
ions. The material remains in a net metallic regime, as visualized in Fig.~\ref{fig:fm-sro/gto}a
when plotting the total spectral function. Both, QP as well as incoherent lower Hubbard peak
are spin polarized, suggesting a mixed behavior of localized and itinerant ferromagnetism. Second,
though initialized as coherent FM, inspecting the real-space ordering interestingly reveals that
the system displays at convergence a ferrimagnetic ordering (see Fig.~\ref{fig:fm-sro/gto}b,c).
The FM metallic layers with Ti1-2 and the FM Mott-insulating layers with Ti3-5 couple 
antiferromagnetically, whereby the PM characterization via regimes I-III remains intact in the 
spin-polarized case. Within the FM orbital-ordered Mott regime the Ti3-4 ions also show nearly 
complete spin polarization. But the most-distant Ti5 ion surprisingly carries a smaller magnetic 
moment. Exchange coupling to the metallic FM layers thus seems to increase the ordered moment. 
These layers in regime II/III are weaker spin polarized, with dominant $d_{xy}$ weight nearest to 
the SrO doping layer. 

The ferromagnetism in the Mott regime is reminiscent of the corresponding ordering of bulk
GTO~\cite{kom07}. In the case of the itinerant TiO$_2$ layers, a combined intra- and inter-layer 
double-exchange-like mechanism~\cite{zen51,and55} may contribute to FM tendencies in the 
interface region with different structural distortions. For a diverse exchange mechanism therein 
also speaks the AFM coupling of Ti to 
the apical oxygens (cf. Fig.~\ref{fig:fm-sro/gto}b), whereas in the Mott layers these ligands appear
FM coupled to the titanium ions. The reason for the AFM coupling of both parts appears more subtle. 
On the basis of the Goodenough-Kanamori rules~\cite{goo63,kan59}, it could be related to the 
'more 180$^{\circ}$-like' orientation of the nearby Ti2-3 $\ket{1}$ orbitals, evident from the 
weaker rotation in the $bc$-plane on Ti2 compared to Ti4 in Fig.~\ref{fig:fm-sro/gto}b. The 
total-energy competition between the paramagnetic and ferrimagnetic state is rather tight, 
favoring the magnetic order by about 10 meV/atom.

\section{Summary and Discussion}
Starting from the Mott-insulating bulk compounds LaTiO$_3$ and GdTiO$_3$ we investigated the
rich and complex electronic structure originating from $\delta$-doping with an SrO monolayer
in a superlattice architecture along the [001] direction. The realistic multi-orbital setup with 
10 TiO$_2$ layers inbetween the doping layers allows to differentiate various correlation regimes
with distance to SrO by the elaborate DFT+DMFT method. An orbital-selective itinerant state at the 
interface changes into an itinerant orbital-polarized doped-Mott state, which transforms to an 
orbital-ordered Mott-insulating region at larger distance. Notably that Mott regime displays even 
stronger orbital polarization than the bulk-insulating case. The transport within the itinerant 
layers is of mixed Fermi-liquid-like- and incoherent character. Generally the $\delta$-doped
GTO case promotes somewhat stronger correlation effects than $\delta$-doped LTO above the
respective ordering temperatures.

Our study reveals the intricate interplay between structural distortions and the electrons' charge, 
orbital, spin degrees of freedom in the heterostructure. The layer- and temperature-dependent 
transitions between the encountered electronic phases pose novel challenging problems in strongly 
correlated electron systems. For instance, further theoretical investigations are necessary to shed 
light on the dependence on the number of SrO layers. Concerning comparison with experiment, our
concrete work serves as a theoretical prediction, since to our knowledge the here here studied 
canonical [100], [010] interface geometry for $\delta$-doped titanates has so far not been 
experimentally investigated. There are recent experimental works on [110], [001] 
interfaces~\cite{zha13,oue13,zha14} that find an overall insulating state for $\delta$-GTO. This
different finding may indeed be traced back to the differences in the interface geometries. Whereas
the ([100],[010]) interface is nearly perfectly flat in our calculation, the ([110],[001]) one
shows some buckling~\cite{zha13}, which we can confirm~\cite{lecto}. Note that due to the different
orientations, the square-like interface Ti sublattice has much stronger rectangular
distortion in the ([110],[001]) geometry. This could more easily trigger the Ti$^{3+}$/Ti$^{4+}$-like 
charge ordering as one option to form the insulating interface state. Furthermore the 
distinct orbital-ordering in Mott-insulating titanates is not easily commensurate across the
([110],[001]) geometry, which might lead to a symmetry-breaking of the original GTO orbital ordering 
based on the $\ket{1}$ state aligned predominantly along the $a$-axis.

Allowing for spin ordering results in a further enrichment of the already sophisticated electronic
structure. A ferrimagnetic ordering, again with itinerant and Mott-insulating regimes, settles in
$\delta$-doped GTO. It results from the subtle exchange-interaction variations due to the
differences in structural distortions, orbital occupations as well as degree of itinerancy.

The present work renders it obvious that the electronic structure characteristics 
in oxide heterostructures are in principle likely to cover the full plethora of many-body 
condensed matter physics within a single (designed) compound. From another perspective, this 
generates vast room for engineering and creating novel states of matter.

\ack
We are grateful to S. Stemmer and L. Balents for helpful discussions. 
Calculations were performed at the Juropa Cluster of the J\"ulich Supercomputing Centre 
(JSC) under project number hhh08. This research was supported by the DFG-FOR1346 project.

\section*{References}
\bibliographystyle{iopart-num}
\bibliography{bibextra}

\end{document}